\documentclass[a4paper,11pt]{article}
\pdfoutput=1 

\usepackage{jheppub} 

\usepackage[T1]{fontenc} 

\usepackage{graphicx}
\usepackage{amssymb}
\usepackage{color}
\usepackage{enumerate}
\usepackage{amsmath}
\usepackage{graphics}

\newcommand{\be}{\begin{eqnarray}}
\newcommand{\ee}{\end{eqnarray}}

\newcommand{\bea}{\begin{eqnarray}}
\newcommand{\eea}{\end{eqnarray}}

\begin{document}

\title{
Unattainability of the Trans-Planckian regime in Nonlocal  Quantum Gravity}

\author[a,b,1]{Fabio Briscese,\note{Corresponding author.}}
\author[c]{Leonardo Modesto}


\affiliation[a]{Academy for Advanced Interdisciplinary Studies, Southern University of Science
	and Technology, Shenzhen 518055, China}
\affiliation[b]{INdAM,  Citt\`{a} Universitaria, P.le A. Moro 5, 00185 Rome, Italy}
\affiliation[c]{Department of Physics, Southern University of Science and Technology, Shenzhen 518055, China}

\emailAdd{briscesef@sustech.edu.cn, briscese.phys@gmail.com}
\emailAdd{lmodesto@sustech.edu.cn}

\abstract{
Based on the ultraviolet asymptotic freedom  of nonlocal quantum gravity, we show that the trans-Planckian energy regime is unattainable in laboratory experiments.  As physical implications, it turns out that the violation of causality, typical of nonlocal field theories, can never be detected in particle accelerators, while the asymptotic freedom of the theory provides an elegant solution to the so called trans-Planckian cosmological problem.
}


\arxivnumber{1912.01878 [hep-th]}

\maketitle

\section{Introduction}

The idea that a renormalizable quantum theory of gravitation can be achieved by introducing higher derivatives or nonlocal interactions was suggested long time ago by Stelle \cite{Stelle}, Krasnikov  \cite{Krasnikov}, and Kuz'min \cite{Kuzmin}. This proposal for an ultraviolet completion of Einstein-Hilbert (EH) gravity was subsequently abandoned, since the specific models studied in \cite{Stelle,Krasnikov,Kuzmin} contain ineradicable ghosts. However, nonlocal  models have been revisited in recent years \cite{Modesto,Review,ModestoLeslaw,briscese1,briscese2,Buoninfante,Buoninfante2},  when it  became clear that, under certain conditions that restrict the type of nonlocality, it is possible to avoid ghosts. In facts,  it has been shown that the complex scattering amplitudes in nonlocal field theories satisfy the Cutkosky rules  \cite{unitarity1,unitarity2,unitarity3}, so that the unitarity is preserved at any pertutbative order in the loop expansion. We mention that higher derivatives and  Lee-Wick quantum gravity are also under current investigation, see \cite{HigherDG,Modesto:2015ozb,Modesto:2016ofr,shapiromodesto,LWqg,anselmi,mannheim}. In particular, Lee-Wick theory has extra complex conjugate poles corresponding to ghosts, that can be consistently removed from the physical spectrum and never go on shell \cite{anselmi}; see also \cite{mannheim} for a discussion of ghost-related issues in fourth-order quantum gravity.

Therefore, in order to achieve a renormalizable quantum theory of gravitational interactions, one is forced to introduce a new nonlocal  or higher derivative action principle. In this paper we will focus on nonlocal quantum gravity \cite{Modesto,Review,ModestoLeslaw,briscese1,briscese2,Buoninfante,Buoninfante2}, but most of the results reported here are still valid for other higher-derivative theories \cite{HigherDG,Modesto:2015ozb,Modesto:2016ofr,shapiromodesto,LWqg,anselmi,mannheim}. 

Nonlocal quantum gravity is   well defined  at classical as well as at quantum level. Indeed, all the classical solutions of the EH theory  are also solution in nonlocal quantum gravity \cite{Yao-dong}, and, most importantly, the stability analysis of such solutions in the nonlocal theory is  the same as in   EH gravity \cite{Stability1,Stability2,nonlocaldesitter}. 
In particular, it has been shown that the Minkowski spacetime is stable under any Strongly Asymptotically Flat (SAF) initial data set satisfying a Global Smallness Assumption (GSA) \cite{Stability1}, as in general relativity. 
Moreover, the model has a satisfactory Starobinsky-like inflation \cite{inflation1,inflation2}, and the spectrum of scalar perturbations generated during inflation is the same as in the local Starobinsky $R^2$ inflation \cite{Starobinsky}, while the spectrum of tensor perturbations is affected by the nonlocality, see \cite{inflation3,inflation4,inflation5} for the details. Indeed, nonlocal  gravity can be falsified by future  measurements of tensor perturbations. At quantum level, the theory is tree-level indistinguishable from EH gravity, namely all the $n-$points scattering amplitudes are the same as in the EH theory \cite{dona}, but it turns out to be super-renormalizable or even finite at higher orders in the loop expansion \cite{Kuzmin,Modesto,ModestoLeslaw,Review}.  Finally, the macroscopic causality based on the Shapiro's time delay is  satisfied \cite{causality}, just because  the tree-level scattering amplitudes  are the same as those of the EH theory. 


In this paper we introduce the new concept of {\em experimental  unattainability} of the trans-Planckian regime in the framework of nonlocal quantum gravity, that is a consequence of the ultraviolet asymptotic freedom  of   nonlocal  fields. In facts, provided that all the fields in the gravitational and the Standard Model sectors are nonlocal, all the particles become asymptotically free
at energies above  $E_{\rm NL}= \ell^{-1}$ \cite{FT1,FT2,FT3,FT4,FT5,FT6,FT8,shapiro3,FT10,barvinsky vilkovisky,barvinsky vilkovisky2,efimov1,efimov2,efimov3,efimovscalar,Universally1,Universally2,Universally3,Universally4}, where $\ell$ is a parameter with dimension of length that fixes the nonlocality scale. As a result,  it is impossible to accelerate particles in the laboratory at energies above  $E_{\rm NL}$ \footnote{We stress that, when we write ``in the laboratory'', we have in mind particle accelerators.}. In facts, at such high energies interactions are very suppressed, and the particles  decouple from any device that could accelerate them. Therefore, provided that $E_{NL} \lesssim E_{P}\equiv \ell_{\rm P}^{-1}$, where $\ell_P=\sqrt{\hbar G/c^3}$ is the Planck length \footnote{In the context of this paper, the nonlocality and Planck scales can be identified setting $E_{NL} \equiv E_{P}$. However, in some nonlocal models, one needs a value $E_{NL} \sim 10^{-5} E_{P}$ in order to fit cosmological data \cite{inflation1,inflation2,inflation3,inflation4,inflation5}.}, we conclude that trans-Planckian energies can not be attained in any laboratory experiment.

As related issue, we  address the problem of causality violations, which are typically expected to occur in nonlocal theories at time scales $\Delta t \sim \ell$.
In order to detect such effect, one should be able to measure time intervals with an accuracy $\ll \ell$ or, equivalently,  to probe the spacetime at a  scale $\Delta x\ll \ell$. Therefore, one should use wave packets much tighter than $\ell$, containing frequencies much higher than $\ell^{-1} = E_{\rm NL}$. However, due to the ultraviolet asymptotic freedom of the model,  particles can not be accelerated above $E_{\rm NL}$, indeed,   such tight  wave-packets can not be produced in particle accelerators. Hence, we conclude that the asymptotic freedom of the theory prevents from the detection of causality violations in laboratory experiments. 



Finally, we show that the ultraviolet asymptotic freedom of nonlocal quantum gravity also provides an elegant solution to the  cosmological trans-Planckian problem \cite{TransPlanckianProblem1,TransPlanckianProblem2,TransPlanckianProblem3,TransPlanckianProblem4,TransPlanckianProblem5,TransPlanckianProblem6,TransPlanckianProblem8}. In facts, according to the inflationary paradigm, cosmological perturbations are seeded by quantum fluctuations of the inflaton field during inflation. At the first stages of inflation, the typical wavelengths of these primordial perturbations are smaller than the Planck length $\ell_{\rm P}$. Indeed one expects that quantum gravity effects should be relevant 	for the evolution of primordial inhomogeneities, so that one should be able to find their footprints in cosmological observations. However, CMB data suggest the contrary \cite{wmap}, showing an agreement with the standard picture based on general relativity coupled to a  scalar field description of matter. This fact is commonly known as cosmological trans-Planckian problem, and the issue of showing why Planck-scale corrections to general relativity are negligible during inflation is still open. We will show that, since nonlocal quantum gravity is Lorentz invariant and it does not contain extra particles, and since all the fields are asymptotically above $E_{NL}$, quantum gravity corrections are naturally suppressed during all the stages of inflation.

We emphasize that the nonlocality explicitly considered in this paper is \textit{de facto} hidden in other quantum  gravity models. For instance, in string theory, nonlocal vertexes of the form $\exp[\Box \, \ell^2]$ appears in the string interaction \cite{string nonlocality1,string nonlocality2,string nonlocality3,string nonlocality4,string nonlocality5,string nonlocality6}. In this case $\ell$ is a string scale that fixes the effective nonlocality scale. Further indications of the emergence of nonlocality at the Planck scale come from  non-commutative theories \cite{amelino,amelino1}, loop quantum gravity \cite{loop}, asymptotic safety
\cite{asymptotic safety}, and causal sets \cite{causal sets}. Moreover, the trace anomaly induced by quantum  corrections due to conformal fields, that is at the basis of the Starobinsky model \cite{Starobinsky}, also induces unavoidable nonlocal terms in the effective action \cite{anomaly1,anomaly2}. In addition, 
we mention that the emergence of hidden nonlocality in quantum gravity and the corresponding impossibility of probing the spacetime below the Planck-length scale has been recently discussed in \cite{addazi}, considering the effect of black holes production in scattering processes.

This paper is organized as follows: in section \ref{section asymptotic freedom} we will review the asymptotic freedom of nonlocal quantum gravity. In section \ref{section unattainalbility and causality violation} we will discuss the implications of this asymptotic freedom, arguing that it makes impossible to accelerate particles up to energies above $E_{NL}$, and, consequently, it makes impossible to detect causality
violations in the laboratory. Finally, in section \ref{section trans-planck} we will discuss the solution of the trans-Planckian problem in the context of nonlocal quantum gravity, and in section \ref{section conclusions} we will briefly summarize our results.

\section{Ultraviolet asymptotic freedom of nonlocal quantum gravity}\label{section asymptotic freedom}




The ultraviolet asymptotic freedom of  higher derivative and nonlocal gravity and gauge theories has been extensively studied in a series of papers \cite{FT1,FT2,FT3,FT4,FT5,FT6,FT8,shapiro3,FT10}. This result has been obtained by means of covariant methods for the calculation of
the effective action in higher derivative quantum field theories and quantum gravity due to Barvinsky and Vilkovisky \cite{barvinsky vilkovisky}, see also \cite{barvinsky vilkovisky2} for review. In this framework, the beta-functions are obtained by perturbative one-loop calculations of the counterterms, and the asymptotic freedom is determined solving the corresponding renormalization group equations.
Since this analysis applies straightforwardly to the nonloacl quantum gravity scenario discussed in this paper, we will skip the detailed calculation of the beta-functions here, remanding the reader to the mentioned literature. In what follows, we will focus on the ultraviolet  behaviour of the coupling constants, and the consequent asymptotic freedom of the theory.

The minimal action for nonlocal quantum gravity reads,
\be
S_{\rm NL} = - \frac{2}{\kappa^2} \int d^4 x \sqrt{-g} \left(R+ \Lambda_c + G_{\mu\nu} \frac{e^{H(\ell^2\Box)} - 1}{\Box} R^{\mu\nu}  \right) \,  ,
\label{NLS}
\ee
where $\ell$  is a parameter with dimensions of a length that fixes  the nonlocality scale, $\Box$ is the covariant d'Alembert operator in curved spacetime, $\kappa^2 = 32 \pi G_N$, $G_{\mu\nu}$ is the Einstein's tensor, $\Lambda_c$ is the cosmological constant, and $\exp H(z)$ is an entire analytic function, that is properly constructed in order to make the theory renormalizable and unitarity (the reader can find more details in \cite{Modesto,Review,ModestoLeslaw}).

Expanding the exponential form factor $\exp H(\ell^2 \Box)$, we recast the action (\ref{NLS}) in the following form  \cite{shapiro3,FT10},

\be
S = - \frac{2}{\kappa^2} \int d^4x \sqrt{-g} \Big\{ \omega_{-2} - \omega_{-1} R    +  \sum_{n=0}^{\infty} \left[ \omega_n^{(0)} R \left(\ell^2\Box\right)^n R + \omega_n^{(2)} R_{\mu\nu} \left(\ell^2\Box\right)^n R^{\mu\nu} \right]
\Big\} \, , 
\label{shapiroAction}
\ee
where the mass dimensions of the parameters are $[\omega_{-2}]=2$, $[\omega_{-1}]=0$, and \mbox{$[\omega^{(0)}_{n}]=[\omega^{(2)}_{n}]= -2$.} 
We can expand the action (\ref{shapiroAction}) in powers of the graviton field around the Minkowski background setting $g_{\mu\nu}= \eta_{\mu\nu}+h_{\mu\nu}$, where $\eta_{\mu\nu}$ is the Minkowski tensor, so that 


\be
S &=& \label{shapiroAgravitons}
\int d^4x  \Big\{  \omega_{-2} \left( 1+ h + h^2 + h^3 + O(h^4) \right)  
- \omega_{-1}  (h \Box h + h^2 \Box h + O(h^4) )  \\
&&
+  \sum_{n=0}^{\infty} \ell^{2n} \left[ \omega_n^{(0)} \left( h \Box^{n+2} h + h^2 \Box^{n+2} h +O(h^4) \right) 
+  \omega_n^{(2)} \left( h \Box^{n+2} h + h^2 \Box^{n+2} h + O(h^4) \right)  \right]    \Big\}  , 
\nonumber 
\ee
where the d'Alembert operator $\Box$ is evaluated on the Minkowski metric, and where we have missed the tensorial structure and all the indices in favour of the explicit structure of the vertices, which do matter in proving the asymptotic freedom. 


According to \cite{FT1,FT2,FT3,FT4,FT5,FT6,FT8,shapiro3,FT10}, the only running couplings are 
\be
\alpha_i \in \left\{ \omega_{-2}, \omega_{-1},  \omega_1^{(0)},  \omega_2^{(0)} \right\} \, , 
\ee
with a trivial running
\be
\alpha_i = \alpha_{i, 0} + \beta_i t \, ,
\ee
where the $\beta_i$ are the beta-functions and $t = \log \mu/\mu_0$, having set $\mu_0 \equiv E_{\rm NL}$. 
Hence, by the following rescaling of the graviton field,
\be
h_{\mu\nu} \rightarrow \alpha_2(t)^{-1/2}  \, h_{\mu\nu}  
\equiv f(t) \, h_{\mu\nu} \, ,
\ee 
where we have defined 
\be \quad f(t)^2 = \frac{f_0^2}{1+ f_0^2 \beta_2 t} \, , 
\ee
the action (\ref{shapiroAgravitons}) turns into 
%

\be
&&  
S    =   \int  d^Dx  \Big\{ \omega_{-2} (1+ f h + f^2 h^2 + f^3 h^3 )) 
-  \omega_{-1} ( f^2 h \Box h + f^3 h^2 \Box h ) 
\\
&&
+ \sum_{n=0}^{\infty} \ell^{2n} \left[  \omega_n^{(0)} \left( f^2 h \Box^{n+2} h 
+ f^3 h^2 \Box^{n+2} h \right)  +\omega_n^{(2)} \left( f^2 h \Box^{n+2} h + f^3 h^2 \Box^{n+2} h \right)\right] + O(f^4 h^4)    \Big\} \nonumber  .
\label{shapiroAgravitonsR}
\ee
At any fixed order in the number of derivatives, the leading nonlinear terms in (\ref{shapiroAgravitonsR}) are those  quadratic in $h$, so that 
\be
&&
 S = \int \! d^4x  \Big\{ \omega_{-2} (1+ f h + f^2 h^2 ) - \omega_{-1}  f^2 h \Box h  
  \nonumber \\ 
&&
+  \sum_{n=0}^{\infty} \ell^{2n} \left[ \omega_n^{(0)} f^2 h \Box^{n+2} h \! 
+ \omega_n^{(2)}  f^2 h \Box^{n+2} h  \right]  + O(f^3 h^3 ) \Big\}   . 
\label{shapiroAgravitonsKin}
\ee
From (\ref{shapiroAgravitonsKin}) it is evident that in the ultraviolet limit all the interactions become negligible, so that the theory is asymptotically free.  We also emphasize that the asymptotic behaviour of the parameter $f(t)$ ensures the validity of the perturbative expansion in powers of $h$ around the Minkowski background.

%

Finally, we can resum all the higher derivative terms in (\ref{shapiroAgravitonsKin}) in order to reconstruct the analytic form factor for the kinetic operator of the graviton field. So far, we obtain the following asymptotic (non-interacting) action in the ultraviolet regime \cite{Review}
\be
&& 
S^{(2)}_{\rm NL} = - \frac{2}{\kappa^2} \int \! d^Dx  \Big[ ( \sqrt{- g}  R)^{(2)} +
G_{\mu\nu}^{(1)} \frac{e^{H(\ell^2 \Box)} -1}{\Box} R^{(1)\mu\nu} 
\nonumber 
\\
&&  + \omega_0^{(0)}(t) (R^{(1)})^2  +  \omega_0^{(2)}(t) R_{\mu\nu}^{(1)} R^{(1) \mu\nu} \Big]
 \, ,
\label{NLlinear}
\ee
where the labels $(1)$ and $(2)$ refer to expansions up to terms linear and quadratic in the graviton respectively.
%

For completeness, we write the linearized equations of motion for the graviton field $h$ as given by the action (\ref{NLlinear}), that read \cite{Stability1,Stability2}
\be
e^{H(\ell^2\Box)} \Box \, h_{\mu\nu}  
= 0 \, .
\label{LEOM}
\ee
We stress that equations (\ref{LEOM}) have the same solutions of the linearized EH theory, because $\exp{H(\ell^2\Box)}$ is an invertible operator by construction \cite{Stability1,Stability2}. This also implies that nonlocal quantum gravity has the same degrees of freedom of general relativity \cite{Stability1,Stability2}, namely the two polarizations of the graviton. The absence of extra degrees of freedom is also at the basis of the unitarity of the theory \cite{unitarity1,unitarity2,unitarity3}, since it prevents the emergence of ghosts.

\section{Unattainability of the trans-Planckian regime and undetectability of causality violations}\label{section unattainalbility and causality violation}

The non locality of the gravitational field introduced in the previous section, including the consequent super-renormalizability and the ultraviolet asymptotic freedom, can be extended to the inflaton field and to the whole sector of the standard model particles, see for instance \cite{efimov1,efimov2,efimov3,Universally1,Universally2,Universally3,Universally4,efimovscalar}. Hereafter, when we will mention nonlocal quantum gravity,  we will refer to this generalized nonlocal framework, in which all the fields are nonlocal and  asymptotically free in the ultraviolet regime.

As a consequence, it results impossible to accelerate any particle to trans-Planckian energies in any laboratory experiment. In facts, due to the ultraviolet asymptotic freedom, all the fundamental interactions become negligible above the energy scale $E_{\rm NL}$, and it is plane that one can not accelerate particles that do not interact with the surrounding environment.  Hence, provided that $E_{NL} \lesssim E_{\rm P}$, we conclude that it is impossible to attain and probe the trans-Planckian regime in particle accelerators.

In force of these considerations, we are now ready to discuss the occurrence of causality violations at the  nonlocality scale $\ell$, that is typical of nonlocal field theories. In particular, we will show that the unattainability of trans-Planckian energies makes  causality violations undetectable in any laboratory experiment. 

To explain how causality violations emerge in nonlocal theories, we consider a toy model consisting of a nonlocal scalar field   in Minkowski spacetime  coupled to an external source $J$. The Lagrangian density of the scalar field reads \cite{efimov1,efimov2,efimov3,carone,Buoninfante2,efimovscalar}
\be
\mathcal{L}_{\phi} = -\frac{1}{2}  \phi \, e^{H(- \ell^2 \Box)}\left(\Box +  m^2 \right)\phi  +  \phi \, J \, ,
\label{action}
\ee
The function $\exp[H(z)]$ must be entire, i.e.,  analytic with no poles at finite $z$, so that the unitarity of the theory is guaranteed \cite{unitarity1,unitarity2,unitarity3}. Moreover,  one requires that $\exp[H(z)]\rightarrow\infty$ for $z\rightarrow -\infty$, in order to improve the ultraviolet convergence of the propagator and enforce the super-renormalizability or finiteness of the theory, see \cite{efimovscalar} for the details. According to (\ref{action}), the equation of motion of the scalar field is: 

\be
e^{H(- \ell^2 \Box)}\left(\Box +  m^2 \right)\phi(x) = J(x) \, ,
\ee
and its solution reads:
\be\label{solution}
\phi(x) = \phi_0(x) + \int  d^4y \,\, G_R(x-y) \, J(y) \, ,
\ee
where $G_R(x-y)$ is the 
Green function satisfying the following equation, 
\bea
\label{GR}
e^{H(- \ell^2 \Box)}\left(\Box +  m^2 \right) G_R(x) = \delta^{(4)}(x) \, .
\eea
The solution of (\ref{GR}) can be easily written in the Fourier space, namely 
\bea 
\label{GRint}
\hspace{-0.2cm}
G_R(x)= \int \frac{d^4k}{\left(2\pi\right)^4} \frac{e^{-\left[H( \ell^2 k^2)+i k x\right]}}{m^2-k^2} =  e^{-H(- \ell^2 \Box)}  G^0_R(x) ,
\eea
where $G^0_R(x)$ is the retarded Green function of the local Klein-Gordon theory, 
\bea\label{G0Rint}
G^0_R(x)= \int \frac{d^4k}{\left(2\pi\right)^4} \frac{e^{-i k x}}{m^2-k^2}  ,
\eea
that satisfies the condition $G_R(x)=0$ for $ x^0 < 0$. 
Replacing (\ref{GRint}) in (\ref{solution}) and integrating by parts, on has 
\be
\label{solution2}
\hspace{-0.5cm}
\phi(x) = \phi_0(x) + \int  d^4y \,\, G^0_R(x-y) \, e^{-H(- \ell^2 \Box_y)} \, J(y) \, ,
\ee
where $\Box_y$ is calculated deriving with respect to $y$. Note that the support of the effective source $e^{-H(- \ell^2 \Box)}  J$ in (\ref{solution2}) is different from that of $J$, i. e. , the source $J$ is smeared  by the action of 
the operator $\exp\left[{-H(- \ell^2 \Box)}\right]$. 

To better understand this fact, we  consider the case in which $J$ is an impulsive source centred at some point $P=(\tau,\vec{q})$ of the spacetime, namely 
\be\label{impulsive source}
\hspace{-0.5cm}
J(y) = g \, \delta^4(y-P) = g  \, \delta(y^0-\tau) \delta^3(\vec{y}-\vec{q}) , 
\ee
where $g$ is a parameter with mass dimension $[g]= - 1$. As an example, in \cite{carone} it has been studied the case in which ${H(- \ell^2 \Box_y)} =  \ell^4 \, \Box^2$, 
and it has been shown that the  effective source $e^{-  \ell^4 \Box_y^2} \, J(y)$ has a support of size $\sim \ell^4$ around $P$. Therefore, the impulsive source (\ref{impulsive source})  is smeared out by the action of the operator $e^{- H(- \ell^2 \Box)}$ into an effective source  of finite support $\Omega_\ell$ with 4-volume $V_\ell \sim \ell^4$. Since the local retarded Green function is such that $G^0_R(x-y) \propto \theta(x^0-y^0)$, the integral in (\ref{solution2}) is nonzero also for $\tau-\ell < x^0 < \tau$. This implies that the scalar field is affected by the source before $J$ is turned on at the time $\tau$, so that there is a violation of causality occurring at time scale $\Delta t \sim \ell$. This simple example shows in a clear fashion how causality is violated in nonlocal theories. We remand the reader to the literature \cite{efimovcausality1,efimovcausality2,efimovcausality3,efimovcausality4,carone} for a detailed discussion of  the causality violation in the scattering amplitudes.

In general, causality violations are due to the fact that the support $\Omega_\ell$ of the effective  source $e^{-H} J$ differs from the support $\Omega$ of the real physical source $J$. Indeed, $\Omega_\ell$ is obtained from $\Omega$ deforming its frontier $\mathit{F}(\Omega)$   by a displacement  of  order  $\ell$, so that $\ell$ defines the scale of the causality violation. Of course, if the source $J$ is localized in a region $\Omega$ of 4-volume $V  \gg \ell^4$, there will be no substantial difference between $\Omega_\ell$ and $\Omega$, so that the causality violation will be negligible. On the other hand, if $\Omega$ has a 4-volume $V  \lesssim \ell^4$ (for instance, in the example of the impulsive source (\ref{impulsive source}) we have $V = 0$) the difference between $\Omega_\ell$ and $\Omega$ will be appreciable. Thus, in order to produce a significant violation of causality, the source $J$ must be localized in  a region of 4-volume $V \lesssim \ell^4$.

Since  the source $J$ represents the interaction of  the scalar field $\phi$ with other particles, $J$ will be a function of other fields, e.g. $J = e \bar\psi \psi$, where $\psi$  is a spinor field. Therefore, in order for $J$ to have a support of 4-volume $V  \lesssim \ell^4$, the field $\psi$ in the given example must be localized in a region of 4-volume $V  \lesssim \ell^4$, so that it must be arranged in wave-packets of width $\Delta  \lesssim \ell$, containing  frequencies $k^0 \gtrsim \ell^{-1}\equiv E_{NL}$. Hence, if we want the effect of causality violations to be relevant, in such a way that it can be detected,  we need to use test-particles of  energy $E \gtrsim E_{\rm NL}$. However, we have already pointed out that the ultraviolet asymptotic freedom of the theory prevents us to produce particles of such an high energy in particle accelerators.
Therefore, we conclude that the causality violations occurring in nonlocal theories cannot be detected in the laboratory.

\section{Solution of the cosmological trans-Planckian problem in nonlocal quantum gravity}\label{section trans-planck}

In this section we  discuss the cosmological trans-Planckian problem \cite{TransPlanckianProblem1,TransPlanckianProblem2,TransPlanckianProblem3,TransPlanckianProblem4,TransPlanckianProblem5,TransPlanckianProblem6,TransPlanckianProblem8}, and its solution in  the framework of nonlocal quantum gravity. Indeed, we aim to explain why quantum gravity effects seems to be negligible even at the inflationary stage, although they should play an important role in the evolution of primordial fluctuations of the inflaton field.  In facts, the theory of cosmological perturbations based on general relativity and on the existence of a primordial inflaton field is in  agreement with current CMB observations \cite{wmap}, while any Planck-scale correction seems to be ruled out by data, at least at the expected order.

The first type of quantum gravity effects that has been considered in \cite{TransPlanckianProblem1,TransPlanckianProblem2,TransPlanckianProblem3,TransPlanckianProblem4,TransPlanckianProblem5,TransPlanckianProblem6,TransPlanckianProblem8} is due to a Lorentz-breaking deformation of the energy-momentum dispersion relation at Planckian energy scales, see \cite{amelino,amelino1} for review. However, the action (\ref{NLS}) is Lorentz (and diffeomorphism) invariant, indeed such effects are absent in nonlocal quantum gravity.

Another class of corrections comes from interactions and higher derivative terms that appear in the action (\ref{NLS}) and in the nonlocal generalizations of the actions of the Inflaton and Standard Model fields. From the effective field theory point of view, it might be hard to explain why higher derivative operators, at linear and nonlinear level in the perturbations, are negligible even at  energies $E \gtrsim E_{NL}$. 

This puzzle has a simple solution in nonlocal quantum gravity \cite{inflation1,inflation2,inflation3}. In facts, part of the higher derivative terms, namely those quadratic in the graviton in (\ref{shapiroAction}), and those quadratic in the other fields in the corresponding actions, are reabsorbed in the nonlocal propagators of the graviton, inflaton,  and Standard Model particles. For instance, all the terms quadratic in the graviton can be recast as in equations (\ref{NLlinear}-\ref{LEOM}), giving a propagator
\be
D_h(k) \propto \frac{i}{k^2\,e^{H(-\ell^2 k^2)}}
\ee
where we have neglected the tensorial structure of the propagator, since it the same as in EH gravity and it is not essential for our discussion; the reader can refer to \cite{Review} for details. What is important here is that, since the function $e^{H(-\ell^2 k^2)}$ is entire, so that it has no zeros for finite values of its argument, this propagator has the same zeros of the graviton propagator in the  EH theory, so that it does not introduce extra particles. This is also evident from the fact that equation (\ref{LEOM}) has the same solutions of the linearized EH equations, so that nonlocal gravity has the same degrees of freedom of general relativity. Therefore, the   propagators of the nonlocal fields have the same poles as the corresponding local fields \cite{unitarity1,unitarity2,unitarity3}, so that nonlocal quantum gravity has no extra degrees of freedom but the graviton, the inflaton, and the Standard Model particles.  The remaining nonlocal terms consists of nonlocal interactions, and can be treated perturbatively, contributing to the scattering amplitudes of the fields. 
However, we have shown in section \ref{section asymptotic freedom} that  all these terms become negligible in the ultraviolet regime, as all the  fields become asymptotically free above  $E_{NL}$.

Thus, since  nonlocal quantum gravity is Lorentz-invariant, and it does not entail extra degrees of freedom, and  all the particles are asymptotically free above $E_{NL}$, all the Planck-scale induced corrections will be suppressed, so that the equations of cosmological perturbations will be the same as in the standard cosmological model. These considerations show that the cosmological trans-Planckian problem is easily solved in the framework of nonlocal quantum gravity.

\section{Conclusions}\label{section conclusions}

In this paper we have considered the generalized framework of nonlocal quantum gravity, in which all the fields are nonlocal. We have argued that, due to the nonlocality of the theory, all the particles are asymptotically free in the ultraviolet regime, above the energy scale $E_{NL}\equiv\ell^{-1}\lesssim E_P$. This implies that the trans-Planckian regime is unattainable in particle accelerators, since it is impossible to accelerate non interacting particles.
This fact, in turn, implies that  causality violations typical of nonlocal models can not be detected  in laboratory experiments.
In facts, we have seen that causality violations occur on a scale $\Delta t \sim \ell$. Indeed, in order to detect them, one needs to probe the spacetime at a scale $\Delta x \lesssim \ell$, and this entails the use of particles of energies $E \gtrsim E_{\rm NL}$. Since the ultraviolet asymptotic freedom of the model prevents the production of such high energy particles, we conclude that it is impossible to measure causality violations in the laboratory.

Finally, we have shown that  the cosmological trans-Planckian problem has a simple and elegant solution in the framework of nonlocal quantum gravity. In facts, this theory is Lorentz-invariant, it has no extra degrees of freedom, and all the particles are asymptotically free at the Planck scale. This implies that all the quantum-gravity induced corrections are suppressed at the first stages of inflation, so that the evolutions of cosmological perturbations is the same as in the standard cosmological model.
We  emphasize some similarity between this picture and  other scenarios with a trans-Planckian cut-off For instance, some authors have formulated a trans-Planckian 
Censorship Conjecture \cite{transplankiancensorship1,transplankiancensorship2,transplankiancensorship3,transplankiancensorship4,transplankiancensorship5,transplankiancensorship6,transplankiancensorship7}, arguing that modes with a length scale smaller than the Planck length must be shielded from classicalizing, in analogy with the Penrose's argument that solutions of GR with naked singularities cannot arise in a full theory. The advantage of nonlocal quantum gravity is that one  does not need to postulate the decoupling of trans-Planckian modes, since this is a direct consequence of the ultraviolet asymptotic freedom of the model.

\end{document}